\documentclass[epj]{svjour}

\usepackage{graphics}
\graphicspath{{images/}}

\usepackage{mathtools,slashed,amssymb,amsmath, bbm, caption,flushend, soul}

\usepackage[numbers,sort&compress]{natbib}
\usepackage[colorlinks,citecolor=blue,urlcolor=blue,linkcolor=blue]{hyperref}

\begin{document}
\title{Quantum tops at the LHC: from entanglement to Bell inequalities}
\author{Claudio Severi\inst{1} \thanks{e-mail: claudio.severi@postgrad.manchester.ac.uk} \and Cristian Degli Esposti Boschi\inst{3,4} \thanks{e-mail: degliesposti@bo.imm.cnr.it} \and Fabio Maltoni\inst{2,4,5} \thanks{e-mail: fabio.maltoni@unibo.it} \and Maximiliano Sioli\inst{2,4} \thanks{e-mail: maximiliano.sioli@unibo.it} }
\institute{Department of Physics and Astronomy, University of Manchester, Manchester, United Kingdom \label{1}
\and Dipartimento di Fisica e Astronomia, Universit\`a di Bologna, via Irnerio 46, Bologna, Italy \label{2}
\and CNR-IMM, Sezione di Bologna,
via Gobetti 101, 40129, Bologna, Italy\label{3}
\and INFN, Sezione di Bologna, via Irnerio 46, Bologna, Italy \label{4}
\and Centre for Cosmology, Particle Physics and Phenomenology, Université catholique de Louvain, Louvain-la-Neuve, Belgium\label{5}
}

\abstract{
We present the prospects of detecting quantum entanglement and the violation of Bell inequalities in $t\bar t$ events at the LHC. We introduce a unique set of observables suitable for both measurements, and then perform the corresponding analyses using simulated events in the dilepton final state, reconstructing up to the unfolded level. We find that entanglement can be established at better than $5 \sigma$ both at threshold as well as at high $p_T$ already in the LHC Run 2 dataset. On the other hand, only very high-$p_T$ events are sensitive to a violation of Bell inequalities, making it significantly harder to observe experimentally. By employing a sensitive and robust observable, two different unfolding methods and independent statistical approaches,  we conclude that, at variance with previous estimates, testing Bell inequalities will be challenging even in the high luminosity LHC run.
}
\maketitle
%

\section{Introduction}

Quantum Mechanics (QM) predicts that when an entangled pair of particles is created, the two--particle wavefunction retains a non--separable character when they are set apart. In particular, correlations on experimental measurements arise even when the observations are space-like separated. Were QM the emergent explanation of an underlying classical theory, the causal structure imposed by relativity would be violated. The issue can also be solved by postulating QM is incomplete, and additional {\it hidden} degrees of freedom exist. Ultimately, whether or not reality is described by QM is matter of experiment. In 1964, Bell proved \cite{bell} classical theories obey correlation limits, {\it i.e.}, Bell Inequalities (BIs), that QM can violate. In the last decades, several experiments have been performed and all results so far agree with QM predictions. Recently, prospects of using hadron colliders to test BIs at unprecedented scales of order of a TeV have emerged \cite{afik2020, fabbrichesi2021, takubo2021, barr2021}. 
In this article, we study the perspectives of experimentally observing entanglement as well as the violation of BIs using spin correlations in $t \bar t$ pairs produced in proton--proton collisions at the LHC, as first suggested in \cite{afik2020} and \cite{fabbrichesi2021}, respectively. We improve on previous studies in several aspects. First,  we introduce  a single set of observables that allow to measure entanglement as well as to assess  BIs violation. Second, we perform an event simulation up to (fast) detector level, fully reconstruct the final states and then determine the quantum observables after unfolding detector effects. We show that our approach is robust under improved physics simulations, such as off-shell and high-order QCD effects, as well as under different choices in the reconstruction procedure. Finally, we carefully examine the stability of the results against  the unfolding procedure. Our analysis confirms the expectations of Ref. \cite{afik2020} on measuring the entanglement at the LHC already with Run II data, while we find very challenging to convincingly prove BIs violation even in the high-luminosity phase of the LHC. 
The paper is organised as follows. We briefly present the general framework of bipartite spin systems in Sec. \ref{sectwo} and the basic physics features  of top quark pair production in Sec. \ref{secfour}. In Sections \ref{secfive} and \ref{secsix} we introduce the quantum observables and, in the case of BIs violation, compare their sensitivity with other proposals. In Sec. \ref{secmadgraph} we present the details of the data simulation and analysis. Our final results are presented in Sec. \ref{results}, while a discussion of the possible loopholes of the Bell experiment are collected in Sec. \ref{ttloophole}. We draw our conclusions in Sec. \ref{conclusions}.

\section{Entanglement and Bell inequalities}
\label{sectwo}

	The state of a system composed of two subsystems $A$ and $B$ is {\it separable} if its density matrix $\rho$ can be written as:
	\begin{equation}
		\rho = \sum_{k = 1}^n p_k \, \rho_A^k \otimes \rho_B^k \,
		,\label{rhoseparable}
	\end{equation}
	where $\rho_A^k$ and $\rho_B^k$ are quantum states for $A$ and $B$, and the coefficients $p_k$ are non-negative and add to one. The density matrix in Eq.~\eqref{rhoseparable} represents a system with {\it classical} probability $p_k$ of being in the state $\rho_A^k \otimes \rho_B^k$. The state is {\it entangled} when it is not separable. Entanglement is a property of a system described in the framework of QM. A different question is whether measurements on a given system can violate BIs. These can be conveniently phrased in terms of Clauser, Horne, Shimony, and Holt (CHSH) inequality \cite{chsh} which states that measurements $a, a'$ and $b , b'$ on subsystems $A$ and $B$, respectively (with absolute value $\leq 1$) {\it classically} must satisfy:
	\begin{equation}
		\big| \langle ab \rangle - \langle ab' \rangle + \langle a'b \rangle + \langle a'b' \rangle \big| \leq 2. \label{chsh_ineq}
	\end{equation}
 A particularly simple example is provided by two particles with non-zero spin. In this case, the measurements entering the CHSH inequality are their spin projections along four axes, $\hat a, \hat a'$ for particle $A$, and $\hat b , \hat b'$ for particle $B$. This corresponds to the situation which we will be considering in the following. 
 
 In QM, entanglement is a necessary condition for violating BIs. However, it is important to stress that in general, {\it i.e.}, for mixed states, the opposite is not true:
	\begin{equation}
		\text{Entanglement}\; \not\!\!\!\!\implies \parbox{7em}{Violation of BIs} \,. \label{logic}
	\end{equation}
A well-known example of states which can be entangled and yet not violate BI's are the so-called bipartite Werner states \cite{werner1989}.
Let us consider a two spin-{$\frac12$} particle system, whose state is described by the simple density matrix:
	\begin{equation}
		\rho = \frac 1 4 \big( \mathbbm 1 \otimes \mathbbm 1 + \sum_{i = 1}^3 C_{ii} \, \sigma_i \otimes \sigma_i \big)\,, \qquad |C_{ii}|<1 \;. \label{simple_rho}
	\end{equation}
This state gives already a (very good) approximation of the pattern of spin correlations in a $t\bar t$ system at the LHC and displays the features that will be important later. 

The specific case where $C_{xx}=C_{yy}=C_{zz}=-\eta$ with $0< \eta<1$ corresponds to the singlet Werner state, while $C_{xx}=C_{yy}=-C_{zz}=\eta$ and cyclic permutations correspond to a triplet of Werner states (with fidelity $F=\frac{3\eta+1}{4}$). It is known that for Werner states, $\eta>1/3$ implies entanglement, while the CHSH inequality is violated when $\eta>\sqrt{2}/2$.\footnote{Werner states are usually expressed as mixtures of Bell states, $\vert \Phi^\pm \rangle$ and $\vert \Psi^\pm \rangle$. In particular, singlet and triplet Werner state correspond to one-parameter families of states that in the high-fidelity limit $F \to 1$ match pure singlet $\vert \Psi^- \rangle$ or (one of the three) triplet states $\vert \Psi^+ \rangle$, $\vert \Phi^- \rangle$, $\vert \Phi^+ \rangle$, respectively.} 

\begin{figure}[h]
 \centering
 \includegraphics[width=.9\linewidth]{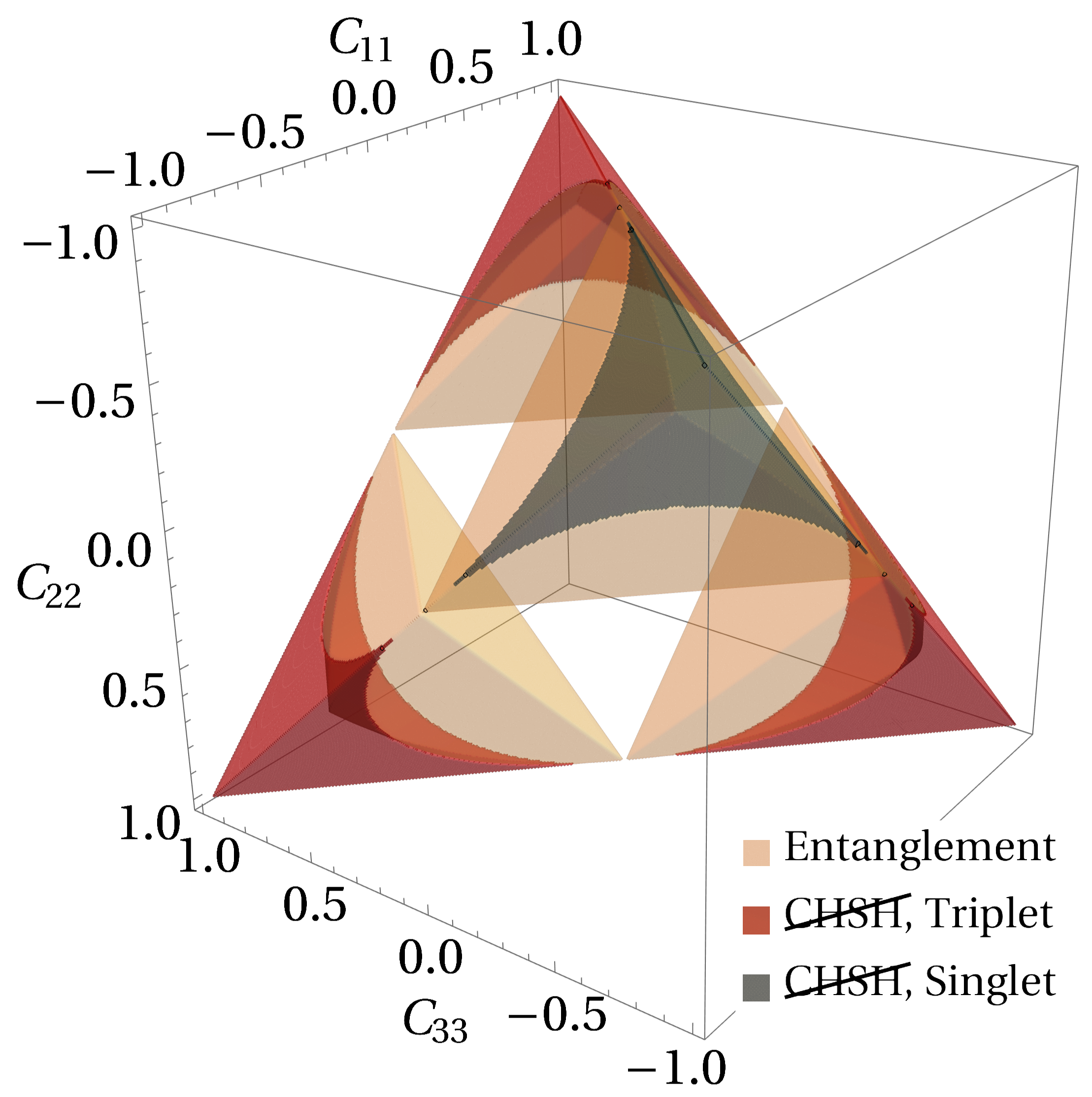}
 \vspace{.5cm}
	\captionsetup{width=\linewidth}
	\caption{Regions of $\{ C_{11}, C_{22}, C_{33} \}$ phase space where the density matrix $\rho$ in Eq.~\eqref{simple_rho} is entangled, and where the CHSH inequality is violated, divided into triplet and singlet. Regions where $\rho$ is not positive definite are not shown.}
	\label{fig:entchsh}
\end{figure}

More in general, one can identify the regions where the state (\ref{simple_rho}) is entangled, using a criterion such as the one proposed in \cite{peres1996}. Independently, one can test whether directions $\hat a, \hat a', \hat b, \hat b'$ exist such that the CHSH inequality, Eq.~ \eqref{chsh_ineq}, is violated, using the theorem proven in \cite{horodecki1995}. The corresponding disjoint regions are depicted in Figure \ref{fig:entchsh}. Entanglement is present in the four light-colored small tetrahedrons which are inscribed in a large tetrahedron, while the CHSH inequality is violated only in the dark-colored sub-regions of the small tetrahedrons. The vertices of the large tetrahedron correspond to pure singlet and triplet Bell states, which we also use to identify the corresponding small tetrahedrons. The closer the $C_{ii}$'s are to vertices of the large tetrahedron the larger amount of spin correlations is present.

\section{Top quark pairs at the LHC} \label{secfour}

\begin{figure}[h]
 \centering
 \includegraphics[width=.5\linewidth]{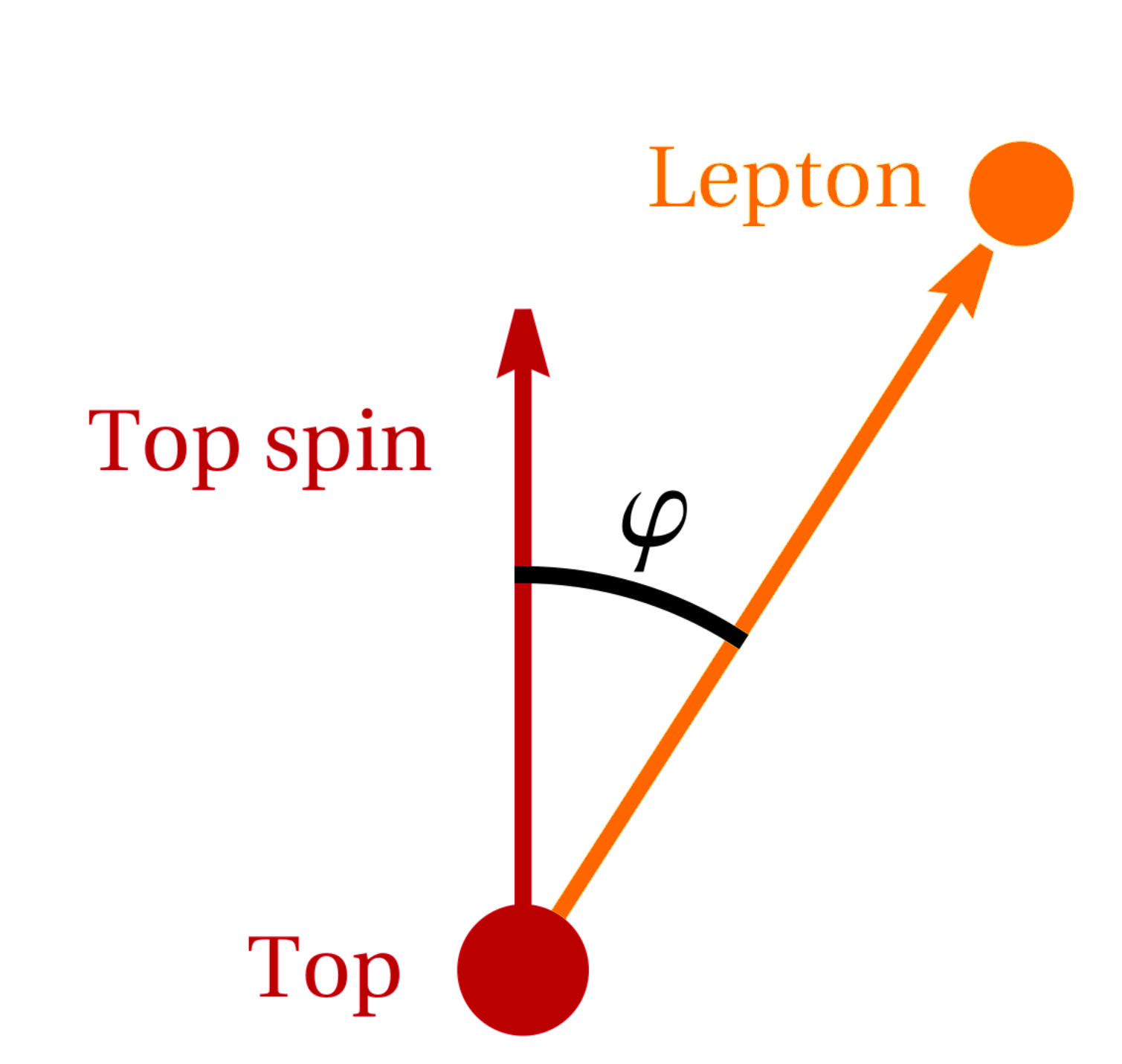}
	\captionsetup{width=\linewidth}
	\caption{ Schematic representation of the decay of a top quark that ultimately leads to the emission of a charged lepton, in the top rest frame.}
	\label{fig:decaydraw}
\end{figure}

\begin{figure}[h]
 \centering
 \includegraphics[width=.7\linewidth]{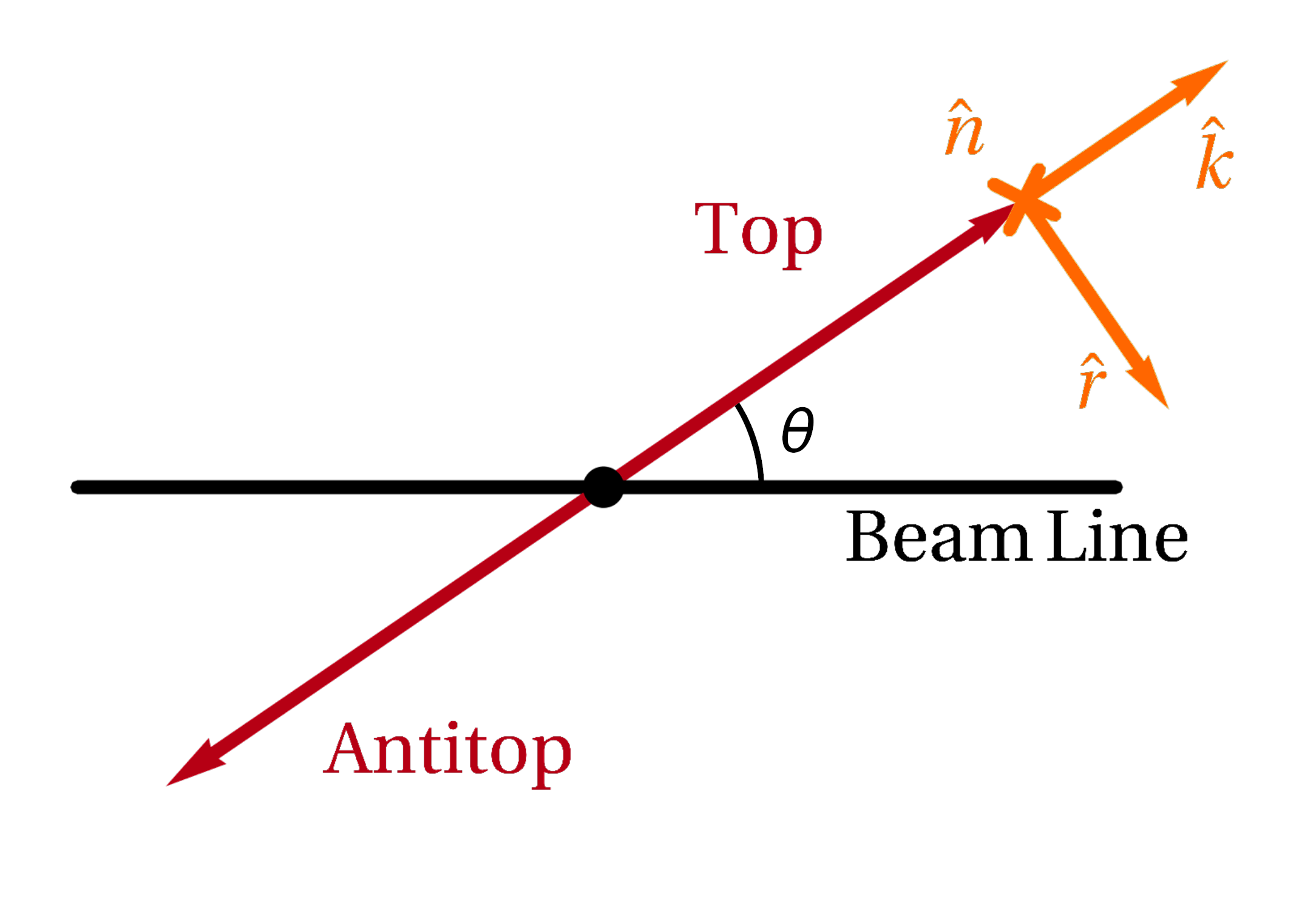}
	\captionsetup{width=\linewidth}
	\caption{ Schematic representation of a $p p \to t \bar t$ event in the center of mass frame, with the helicity basis $\lbrace \hat k, \hat r, \hat n \rbrace$ drawn, together with the scattering angle $\theta$. The $\hat n$ axis is into the page.}
	\label{fig:topdraw}
\end{figure}

	 Top quarks are unique candidates for high energy Bell tests. This follows from three concurring facts. First, since $m_t \, \Lambda^{-2}_{\text{\tiny QCD}} \gg \Gamma_t^{-1}$, top quarks decay semi-weakly before their spin is randomised by chromomagnetic radiation. Second, the leading top quark production mechanism at hadron colliders involves $t \bar t$ pairs where top quarks are not polarised yet their spins are highly (and not-trivially) correlated in different areas of phase space. Third, thanks to left-handed nature of the weak interactions, the charged lepton emerging from the two-step decay $t \to W b$ and $W \to \ell \nu$ turns out to be $100\%$ correlated with the spin of the mother top quark, {\it i.e.} the top quark differential width is given by:
	\begin{equation}
		\frac{1}{\Gamma} \frac{d\Gamma}{d \cos \varphi} = \frac{1 + \alpha \cos \varphi}{2}\,,
	\end{equation}
	with the spin analyzing power $\alpha$ attaining the largest possible values, {\it i.e.,} $\pm 1$. We denote $\varphi$ the angle between the top spin and the direction of the emitted lepton in the top quark rest frame, see Figure \ref{fig:decaydraw}. As a result, the lepton can be considered as a proxy for the spin of the corresponding top quark and the correlations between the leptons as a proxy for those between the top quark spins.
 Assuming no net polarisation is present \footnotemark, the density matrix for the spin of a $t \bar t$ pair can be written as:
\footnotetext{Strong $t\bar t$ production does not lead to polarised top quarks, as parity is conserved \cite{bernreuther2001}. EW effects (and possibly also absorptive parts from loops), on the other hand, can give rise to a net top quark polarisation. However, they have been estimated to be very small \cite{frederix2021}, and therefore are neglected here.}
	\begin{equation}
		\rho = \frac 1 4 \big( \mathbbm 1 \otimes \mathbbm 1 + \sum_{i,j = 1}^3 C_{ij} \, \sigma_i \otimes \sigma_j \big) \label{rho}\,.
	\end{equation}
	where the first term in the tensor product refers to the top and the second term to the anti-top quark. The $C_{ij}$ matrix encodes spin correlations, and it is measurable. Note that Eq.~\eqref{rho}, which will be used in the following, is more general than the simple density matrix in Eq.~\eqref{simple_rho} considered in Section \ref{sectwo}, since $C$ is allowed to have off-diagonal entries. However, since in practice $C_{ij} \approx C_{ji}$, the $C$ matrix can be made (almost) diagonal with an appropriate choice of basis, thus reducing the $t \bar t$ system to Eq.~\eqref{simple_rho}. The differential cross section for $p p \to t \bar t \to \ell^+ \ell^- b \bar b \nu \bar \nu$ can be expressed as \cite{bernreuther2004}:
	\begin{equation}
		\frac{1}{\sigma} \frac{d \sigma}{dx_{ij}} = \frac{C_{ij} \, x_{ij} - 1}{2} \log \big|x_{ij} \, \big|, \label{dcoscos}
	\end{equation}
	where $x_{ij} \equiv \cos \theta_{i} \cos \bar \theta_{j}$, $\theta_i$ is the angle between the antilepton momentum and the $i$-th axis in its parent top rest frame, and $\bar \theta_j$ the angle between the lepton momentum and the $j$-th axis in its parent anti-top rest frame. In particular, Eq.~\eqref{dcoscos} implies:
	\begin{equation}
		-9 \langle x_{ij} \rangle = C_{ij},
	\end{equation}
	a relation that allows direct measurement of the $C$ matrix. Spin is measured fixing a suitable reference frame. An advantageous choice is the {\it helicity basis} $\lbrace \hat k, \hat r, \hat n \rbrace$, 
	\begin{equation}
		\begin{dcases} 
			\ \hat k = \text{top direction} \\ 
			\ \hat r = \frac{\hat p - \hat k \cos \theta}{\sin \theta}\\ 
			\ \hat n = \hat k \times \hat r,
		\end{dcases} \label{helbasis}
	\end{equation}
	 where $\hat p$ is the beam axis and $\theta$ is the top scattering angle in the center of mass frame, see also Figure \ref{fig:topdraw}. The helicity basis is defined in terms of the top quark and also applies to the antitop.\footnote{We follow the sign convention of \cite{afik2020}.} Relevant reference frames are reached in a two step process: a $\hat z$ boost from the laboratory to the $t \bar t$ center of mass frame, then a $\hat k$ boost to each top's rest frame.

	 The amount and type of spin correlations strongly depend on the production mechanism as well as the phase space region (energy and angle) of the top quarks. Two complementary regimes are important: at threshold, {\it i.e.}, when the top quarks are slow in their rest frame, and when they are ultra-relativistic. At threshold, gluon fusion $gg \to t \bar t$ leads to an entangled spin-$0$ state while $q \bar q \to t \bar t$ to a spin-$1$ state. The latter is subdominant at the LHC and acts as an irreducible background \cite{afik2020}. 

\section{Observation of entanglement} 
\label{secfive}

 It can be shown \cite{peres1996} that the $t \bar t$ spin density matrix in Eq.~\eqref{rho} is separable (that is, not entangled) if and only if the partial transpose $(\mathbbm 1 \otimes T) \, \rho$, obtained by acting with the identity on the first term of the tensor product and transposing the second, is positive definite. As shown in \cite{afik2020}, this implies that
	\begin{equation}
		 \big| C_{11} + C_{22} \big| - C_{33} > 1\, \label{ent}
	\end{equation}
	is a sufficient condition for the presence of entanglement. It generalises the Werner condition $\eta>1/3$ to the case where the $C_{ii}$'s are not equal. The inequality in Eq.~\eqref{ent} does not depend on the basis, but we will use the helicity basis in  Eq.~\eqref{helbasis} in the following.
	
	At $t \bar t$ production threshold $C_{kk} + C_{rr} < 0$, so inequality of Eq.~\eqref{ent} reads:
	\begin{equation}
	 - C_{kk} - C_{rr} - C_{nn} > 1. \label{entthr}
	\end{equation}
	The second regime corresponds to high transverse momentum top quarks, {\it i.e.}\@ when the system is characterised by $m_{t \bar t} \gg m_t$ and CMF scattering angle $\theta \sim \frac{\pi}{2}$. In this case, an entangled spin-$1$ state is produced as a consequence of conservation of angular momentum regardless of production channel. Since in this region $C_{kk} + C_{rr} > 0$, inequality in Eq.~\eqref{ent} is written as:
	\begin{equation}
	C_{kk} + C_{rr} - C_{nn} > 1. \label{enthigh}
	\end{equation}

\section{Observation of a violation of the BIs} 
\label{secsix}
	
	Spin correlations at threshold are strong enough to show entanglement, yet not enough to allow the observation of a violation of the BIs. In addition, as we will discuss in the following, top quarks are moving slowly and their decays are usually not causally disconnected. A violation of the BIs can, however, be expected at large $m_{t \bar t}$ and $\theta\simeq \pi/2$. In this region of the phase space mass effects are subdominant. Different strategies to experimentally observe violations of BIs exist, we present them in the following. \smallskip

	The CHSH inequality, Eq.~\eqref{chsh_ineq},  can be written in terms of the $C_{ij}$ matrix as:
	\begin{equation}
		\Big| \sum_{ij} C_{ij} \, ( a_i b_j - a_i b'_j + a'_i b_j + a'_i b'_j ) \Big| \leq 2. \label{chsh_c}
	\end{equation}
 As shown in Refs. \cite{horodecki1995, chen2013} the maximal value predicted by QM in the CHSH inequality of Eq.~\eqref{chsh_ineq} is:
	\begin{equation}
		\max_{a \, a' \, b \, b'} \Big| \sum_{ij} C_{ij} \, ( a_i b_j - a_i b'_j + a'_i b_j + a'_i b'_j ) \Big| = 2\sqrt{\lambda + \lambda'}, \label{eqthm}
	\end{equation}
	where $\lambda$ and $\lambda'$ are the two largest eigenvalues of $C^T C$. 
	The maximal value is obtained for the following choice of directions:
	\begin{align}
		a = Cd, &\quad a' = Cd' \label{axisa}, \\
		b = d \cos \varphi + d' \sin \varphi, &\quad b' = - d \cos \varphi + d' \sin \varphi, \label{axisb}
	\end{align}
	where $d$ and $d'$ are eigenvectors of $C^T C$ corresponding to the eigenvalues $\lambda, \lambda'$, and $\tan \varphi = \sqrt{\frac{\lambda'}{\lambda}}$. \smallskip

 Following Ref. \cite{afik2020} and Eqs. \eqref{axisa} and \eqref{axisb}, it can be shown that in the limit $m_{t \bar t} \gg m_t$ and $\theta = \frac \pi 2$, the axes (in the helicity basis):
	\begin{gather}
	 a = (0, \, 1, \, 0), \quad a' = (0, \, 0, \, 1), \nonumber \\
	 b = (0, \, -\frac{1}{\sqrt{2}}, \, \frac{1}{\sqrt{2}}), \quad b' = (0, \, \frac{1}{\sqrt{2}}, \, \frac{1}{\sqrt{2}})\,, \label{axis}
	\end{gather}
	 correspond to the optimal choice (up to NLO QCD corrections. Adopting the axes in Eq.~\eqref{axis}, the CHSH inequality in Eq.~\eqref{chsh_ineq} can then be cast in a particularly simple form:
	\begin{equation}
		\sqrt 2 \, \big| -C_{rr} + C_{nn} \big| \leq 2, \label{chshhigh}
	\end{equation}
	which, once again, generalises the CHSH condition derived for Werner states. Note the $\hat k$ axis does not appear in Eq.~\eqref{chshhigh}, consistent with the physical argument that in a Bell experiment the spin (helicity) of a massless particle is measured on a plane perpendicular to its motion. A spin correlation experiment in this regime is equivalent to the usual quantum optics experiment with entangled photons, except for being characterised by a $10^{12}$ times larger energy. \smallskip

In general, the optimal choice of directions $\hat a, \hat a', \hat b, \hat b'$ using Eqs. \eqref{axisa} and \eqref{axisb} can be evaluated for each point in phase space, in terms of $m_{t \bar t}$ and $\theta$. In this case, for each $t \bar t$ event, one can determine the optimal choice of axes and evaluate Eq.~\eqref{chsh_c}. This strategy should, in principle, maximise the effects of spin correlations. However, it also enhances the effects of systematic uncertainties in the event reconstruction, i.e., in the assignment of $m_{t \bar t}$ and $\theta$. \smallskip

	As suggested in \cite{fabbrichesi2021}, following from the result in Eq.~\eqref{eqthm}, one can use:
	\begin{equation}
		\lambda + \lambda' \leq 1, \label{chsh_fabbrichesi}
	\end{equation}
	as the CHSH inequality. This strategy entails a rather serious bias. Since each event is used to find the axes maximizing Eq.~\eqref{chsh_ineq} and to evaluate Eq.~\eqref{chsh_ineq} itself, this method automatically selects upwards-shifting statistical fluctuations. In other words, random fluctuations are more likely to drive the eigenvalues of $C^T C$ in the positive direction rather than in the negative one, and furthermore, when selecting the two largest eigenvalues $\lambda$ and $\lambda'$ one is more likely to pick the ones that fluctuated up rather than those that fluctuated down. As a result, the estimated value of $\lambda + \lambda'$ is on average larger than the true value.
	 The amount of bias present in $\lambda + \lambda'$ is a notoriously difficult quantity to evaluate \cite{furedi1981}. To overcome the issue of correcting for the bias, one can use the observable $\lambda + \lambda'$ in an hypothesis test as follows. First, the entries $C_{ij}$ are reconstructed from (simulated) data, producing the matrix $C$. Then, the entries of $C$ are smeared randomly many times according to their uncertainties, and the resulting distribution of $\lambda + \lambda'$ is constructed. The same procedure is applied to the "classical" correlation matrix
	\begin{equation}
		C_{\text{classical}} = \begin{pmatrix} \frac{1}{\sqrt 2} & 0 & 0 \\ 0 & \frac{1}{\sqrt 2} & 0 \\ 0 & 0 & -\frac{1}{\sqrt 2} \end{pmatrix}\,,
	\end{equation}
	which can be considered as the worst--case scenario to reject, as it imitates the spin correlations we want to observe but has $\lambda + \lambda' = 1$. Then, the probability that upon smearing of $C_{\text{classical}}$ one finds:
	\begin{equation}
	\text{Med}[\lambda + \lambda'] > (\lambda + \lambda')_{\text{classical}}
	\end{equation}
	is the significance for rejecting the "classical" hypothesis, given the data. \smallskip
	
\begin{figure}
    \centering
    \includegraphics[width=.8\linewidth]{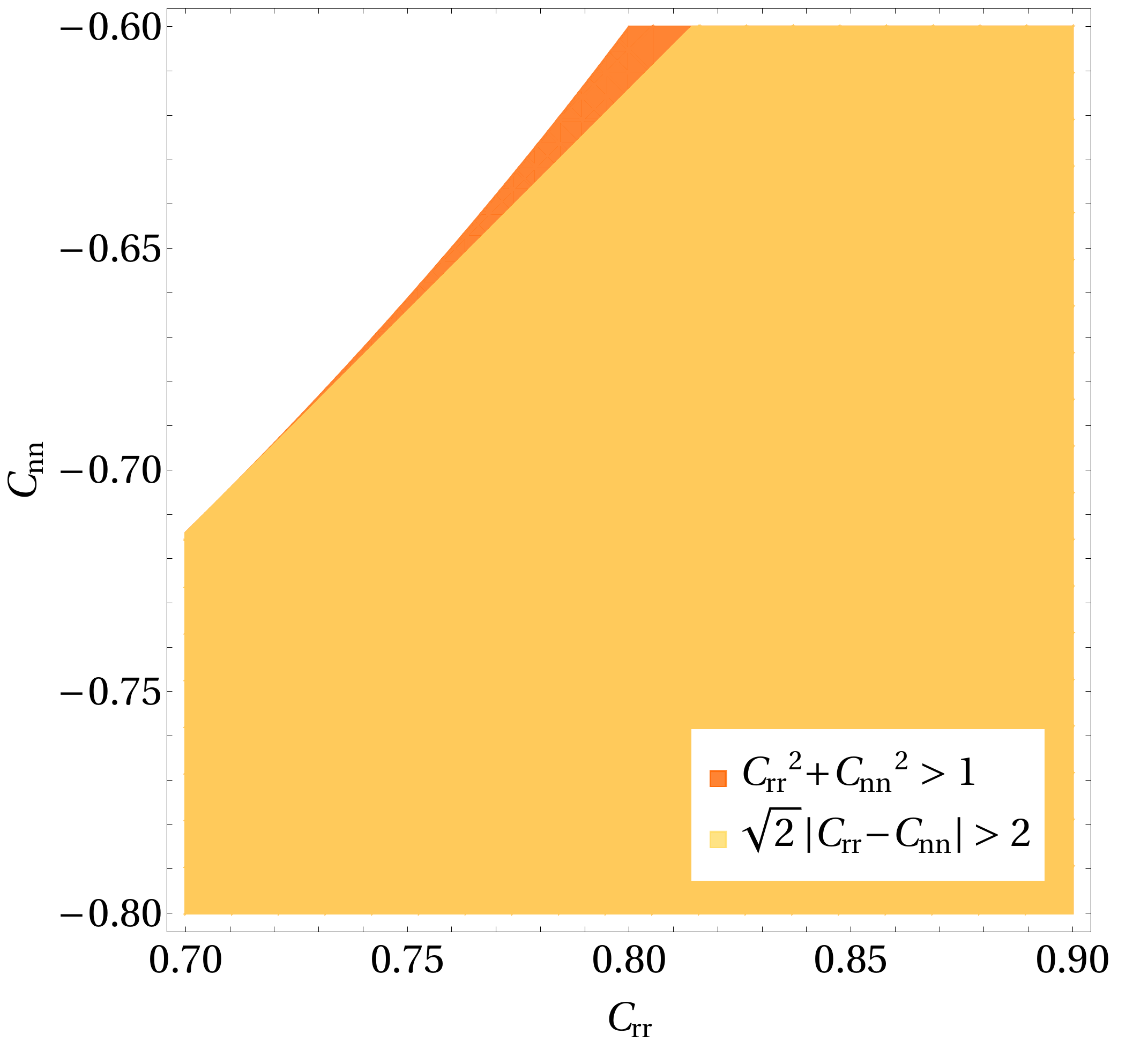}
	\captionsetup{width=\linewidth}
	\caption{Comparison between the conditions in Eqs. \eqref{chshhigh} and \eqref{chsh_fabbrichesi_io} in the $C_{rr} - C_{nn}$ plane, in the region of interest for this analysis.}
	\label{fig:compare}
\end{figure}

	We conclude this section with an observation regarding Eq.~\eqref{chsh_fabbrichesi}.  Assuming the $C_{ij}$ matrix is approximately diagonal, and $|C_{kk}| < |C_{rr}|$,  $|C_{kk}| < |C_{nn}|$, which is satisfied in the signal region, one obtains from \eqref{chsh_fabbrichesi}:
	\begin{equation}
		\lambda + \lambda' \approx C_{rr}^2 + C_{nn}^2 \leq 1 \,. \label{chsh_fabbrichesi_io}
	\end{equation}
	Inequality \eqref{chsh_fabbrichesi_io} identifies a disk and looks different from Eq.~\eqref{chshhigh}, which is linear. However, in the region where the experimental measurement most likely lies, $C_{rr}$ and $C_{nn}$ are constrained to be $0.7 \lesssim C_{rr} \lesssim 0.9$, $-0.8 \lesssim C_{nn} \lesssim -0.6$ and the two conditions are essentially equivalent, see Figure \ref{fig:compare}. This shows that Eq.~\eqref{chshhigh} is as a sensitive indicator for the violation of BIs as the reference-frame independent bound provided by Eq.~\eqref{chsh_fabbrichesi}.
	
\section{Simulation and analysis} 
\label{secmadgraph}
	
 We generate events corresponding to proton--proton collisions at $\sqrt s = 13 \, \text{TeV}$ resulting in a $\ell^- \, \ell^+ \, \nu \, \bar \nu \, b \, \bar b$ final state. Leptons $\ell$ only include electrons and muons, both in same and different flavor combinations. Non--resonant, single resonant and double resonant top quark diagrams are summed, as well as diagrams in which no top quarks exist, yet the final state is two bottom quarks, two leptons, and two neutrinos. Events are generated using {\sf MadGraph5\_aMC@NLO} \cite{madgraph} at Leading Order (LO) within the Standard Model. This amounts to about $5000$ diagrams summed. No kinematic cuts are imposed except for a lower limit of $5 \, \text{GeV}$ on the invariant mass of same flavor lepton pairs, to keep the $\gamma^* \to \ell^+ \ell^-$ splitting infrared safe, and a $10 \, \text{GeV}$ lower limit for charged leptons $p_T$. The rate is normalised to an estimate of the experimental cross section of $p p \to \ell^- \, \ell^+ \, \nu \, \bar \nu \, b \, \bar b$ obtained with recent measurements \cite{pdg}. Hard processes are showered and hadronised using {\sf Pythia 8} \cite{pythia8}. Events are then passed to the {\sf Delphes} framework \cite{delphes} for fast trigger, detector and reconstruction simulation, using the setup of ATLAS detector at the LHC. 
 In our simplified analysis, we only consider irreducible backgrounds, consisting of contributions to the $\ell^- \, \ell^+ \, \nu \, \bar \nu \, b \, \bar b$ final state without an intermediate $t \bar t$ pair, or with an intermediate $t \bar t$ pair that does not decay into prompt light leptons. These contributions become negligible when the kinematics of two on-shell top quarks is correctly reconstructed. Further sources of background include $t \bar t \, V$ events, diboson events, and misidentification of leptons. These backgrounds are known to amount to a few percent of the total \cite{atlas5} and are neglected. The same flavor channel also receives a contamination from $Z + \text{jets}$ events, whose number, after cuts, is at the percent level \cite{cms3}, comparable with other backgrounds already quoted. At the selection level, we require the presence of at least two jets with $p_T > 25 \, \text{GeV}$ and $|\eta| < 2.5$. At least one jet has to be $b$--tagged. If only one jet has been $b$--tagged, we assume the second $b$-jet is the one not $b$--tagged with the largest $p_T$. We require exactly two leptons of opposite charge, both with $p_T > 25 \, \text{GeV}$ and $|\eta| < 2.5$. Both leptons must pass an isolation requirement. In the $e^+ e^-$ and $\mu^+ \mu^-$ channels, $Z$ + jets processes are suppressed by requiring $p^{\text{miss}}_{T} > 40 \,\text{GeV}$ and $20 \, \text{GeV} < m_{\ell^+ \ell^-} < 76 \, \text{GeV}$ or $m_{\ell^+ \ell^-} > 106 \, \text{GeV}$. 
 Before reconstructing neutrinos, the measured momenta of $b$ quarks and the value of the missing energy are smeared according to the simulated distribution of reconstructed values around true values. Neutrinos are then reconstructed solving for the kinematics of $t \bar t \to \ell^- \, \ell^+ \, \nu \, \bar \nu \, b \, \bar b$. The solution is assigned a weight proportional to the likelihood of producing a neutrino of the given reconstructed energy in a $p p \to t \bar t \to \ell^- \, \ell^+ \, \nu \, \bar \nu \, b \, \bar b$ at $\sqrt s = 13 \, \text{TeV}$ in the SM. If many solutions exist for the kinematics, the solution yielding the smallest $m_{t\bar t}$ is considered.
 The smearing on $b$ quarks and $p^{\text{miss}}$ is repeated $100$ times. There is a twofold ambiguity in assigning $b$ quarks to jets, so reconstruction is performed twice for each event. The assignment yielding the largest weight is chosen, and the final $p_\nu$ and $p_{\bar \nu}$ are calculated as a weighted average. Further details on the performance of our event reconstruction algorithm can be found in \cite{iome}. The reconstructed distributions of $\cos \theta_{i} \cos \bar \theta_{j}$ appearing in Eq.~\eqref{dcoscos} are unfolded using the iterative Bayesian method \cite{iterativebayesian} implemented in the {\sf RooUnfold} framework \cite{roounfold}, and the final statistical uncertainty on the $C_{ij}$ matrix is computed taking into account the bin-to-bin correlations introduced by the unfolding process.
 
 As a validation of our unfolding procedure, we have repeated the unfolding varying the number of iterations (from 3 to 6)  and the number of bins (from 6 to 12) and results are found to be stable. The unfolding performance worsens when the number of iterations is increased above $\sim 10$, as expected, since the iterative Bayesian method \cite{iterativebayesian} reproduces the simple inversion of the response matrix without regularisation for $n_{\text{iterations}} \to \infty$. As a further check, we have unfolded our reconstructed distributions with the Singular Value Decomposition technique proposed in \cite{hocker1995}, also implemented in {\sf RooUnfold}, for various number of bins (from 6 to 12) and various choices of the regulating parameter $k$ (from 3 to 5). Results are consistent with the iterative method and stable under change of parameters. When run over $35.9 \, \text{fb}^{-1}$ of simulated luminosity with the same kinematical cuts in \cite{cms3}, our analysis produces statistical uncertainties at the unfolded level that are compatible with those found by the CMS Collaboration. 

In order to verify the robustness of our observable definition and reconstruction method against higher-order QCD effects, we have generated $250 \, \text{fb}^{-1}$ of $p p \to t \bar t$ events at $\sqrt s = 13 \, \text{TeV}$ at Next-to-Leading Order (NLO) in QCD with {\sf MadGraph5\_aMC@NLO}. Since NLO QCD corrections to the $C_{ij}$ matrix are known to be small \cite{bernreuther2001}, top spin correlations and finite width effects have been taken into account using {\sf MadSpin} \cite{madspin}. The test statistics, Eqs.~\eqref{entthr}, \eqref{enthigh}, and \eqref{chshhigh}, are then re-evaluated using the event reconstruction algorithm cited above. Deviations in our region of interest are seen at the percent level, meaning the algorithm is well--behaved under the introduction of NLO QCD corrections, and missing higher-order terms in our LO analysis are sub-leading with respect to statistical uncertainty in realistic LHC scenarios.

\section{Results}
\label{results}

\begin{figure}
 \centering
 \includegraphics[width=.8\linewidth]{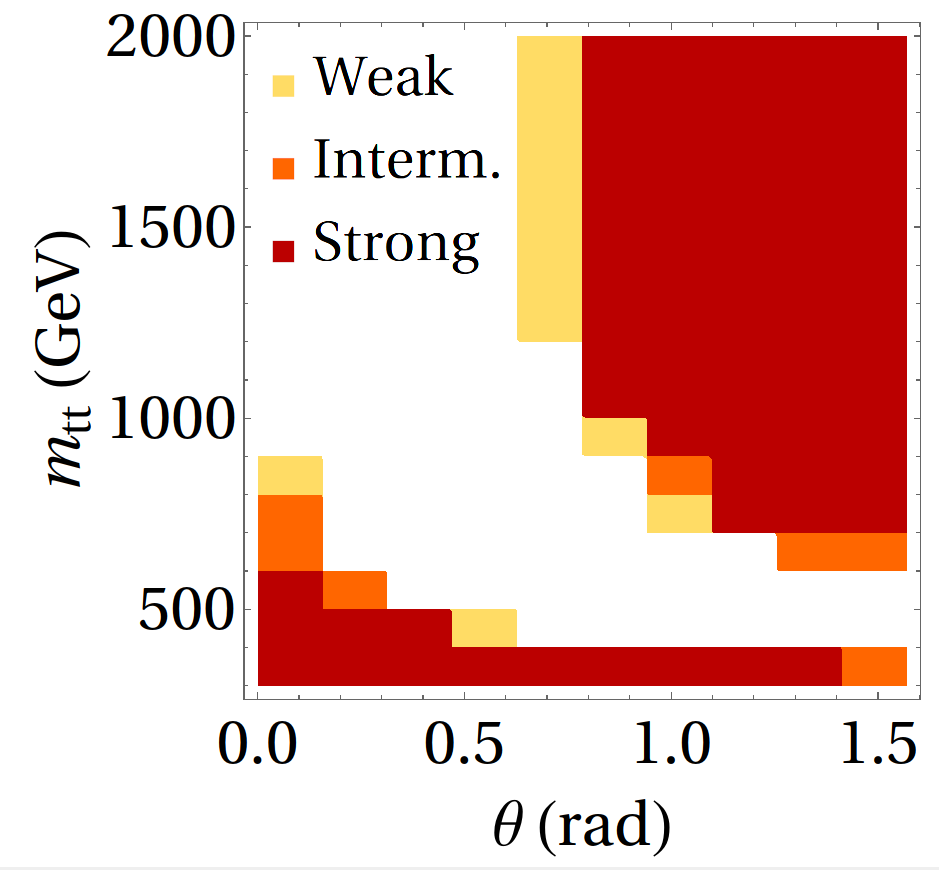}
	\captionsetup{width=\linewidth}
	\caption{Selections in $m_{t \bar t} - \theta$ space for the detection of entanglement.}
	\label{fig:cutsEnt}
\end{figure}

As a first step, we consider the observation of entanglement. The two signal regions of interest are i) at threshold and ii) at large $p_T$. We consider three different selections, characterised by different trade-offs between keeping the largest possible statistics and maximising the correlations. The three selections are shown explicitly in Figure \ref{fig:cutsEnt}, with the ``strong'' selection being completely contained in the ``intermediate'' selection, that in turn is contained in the ``weak'' selection.
Results are collected in Table \ref{table:ent_results}, together with an estimate of the cross section included in each selection. When considering the LHC Run 2 luminosity of $139 \, \text{fb}^{-1}$, the expected statistical significance for the detection of entanglement is of order $5\sigma$ or more in both signal regions. \smallskip

	\begin{table*}
	{
		\centering
		\begin{tabular}{||l l c | c c||} \hline
		 & & & $|C_{kk} + C_{rr}| - C_{nn}$ & Significance for $> 1$ \\
 			 Region & Selection & Cross section & {Reconstructed} & {[$139 \, \text{fb}^{-1}$] } \\ \hline
		 & Weak & $14 \, \text{pb}$ & $1.31 \pm 0.02$ & $> 5 \, \sigma$ \\
 			Threshold& Intermediate & $12 \, \text{pb}$ & $1.34 \pm 0.02$ & $> 5 \, \sigma$ \\
 			& Strong & $10 \, \text{pb}$ & $1.38 \pm 0.02$ & $> 5 \, \sigma$ \\
 			\hline
 		& Weak & $1.9 \, \text{pb}$ & $1.32 \pm 0.06$ & $5 \, \sigma$ \\
 			High-$p_T$ & Intermediate & $1.6 \, \text{pb}$ & $1.36 \pm 0.07$ & $5 \, \sigma$ \\
 		 & Strong & $0.9 \, \text{pb}$ & $1.42 \pm 0.10$ & $4 \, \sigma$ \\
 			\hline
		\end{tabular}
		\captionsetup{width=\linewidth}
		\caption{Results for the entanglement markers in Eqs.~\eqref{enthigh} and \eqref{entthr} in the two signal regions, for the selections explained in the text. For each selection an estimate of the cross section at $13 \, \text{TeV}$ is also reported. The quoted results are the average of several (5) independent simulated experiments using $139 \, \text{fb}^{-1}$ of simulated luminosity (LHC Run 2) each. Uncertainty is statistical only.}
		\label{table:ent_results}
		}
	\end{table*}

\begin{figure}
 \centering
 \includegraphics[width=.8\linewidth]{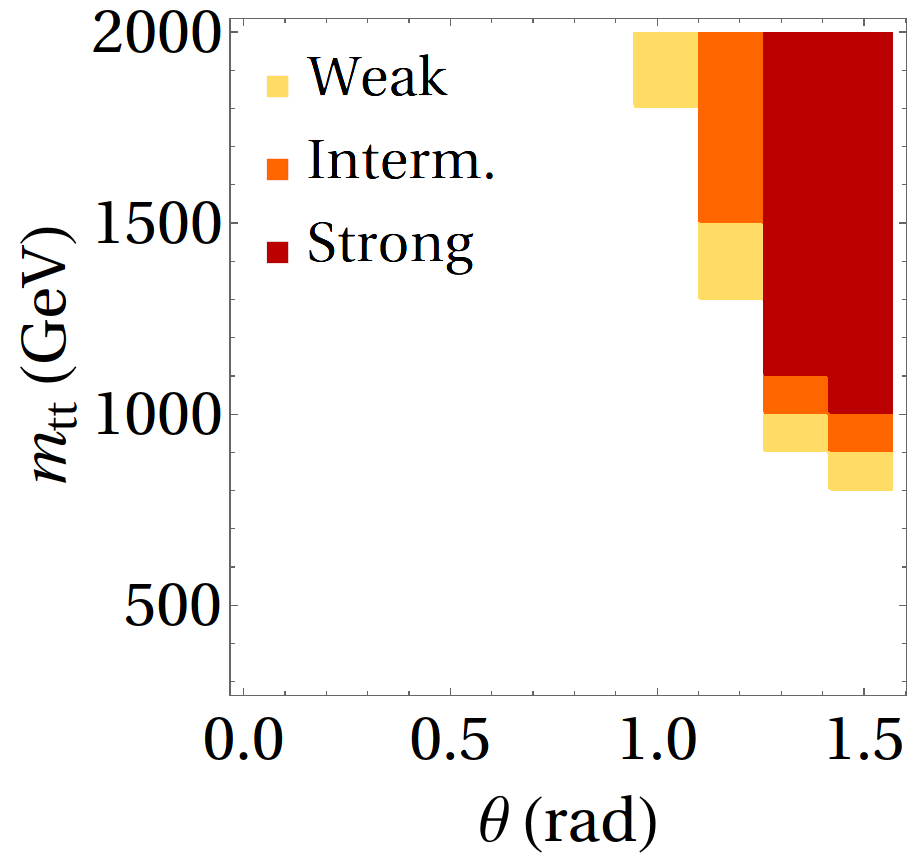}
	\captionsetup{width=\linewidth}
	\caption{Selections in $m_{t \bar t} - \theta$ plane for the observation of a violation of the CHSH inequality. }
	\label{fig:cutsCHSH}
\end{figure}

The strategy to observe a violation of BIs is the same as the one employed for entanglement. In this case, however, we only consider one signal region, corresponding to events with $m_{t\bar t}$ of order TeV and $\theta$ close to $\frac \pi 2$, and move directly to simulating experiments using $350 \, \text{fb}^{-1}$ of luminosity. We consider three selections, shown explicitly in Figure \ref{fig:cutsCHSH}, with the same ``strong''/``intermediate''/``weak'' hierarchy as before. Assuming an average detector efficiency of $12\%$ in successfully reconstructing parton--level $t \bar t$ events, consistent with the results of our simulations, our three different selections should yield approximately $10^4$, $5 \cdot 10^3$, and $3 \cdot 10^3$ events respectively at the end of Run 3 of the LHC, and a factor of $\sim 10$ more after the High--Luminosity Run.

\begin{figure}
 \centering
 \includegraphics[width=.9\linewidth]{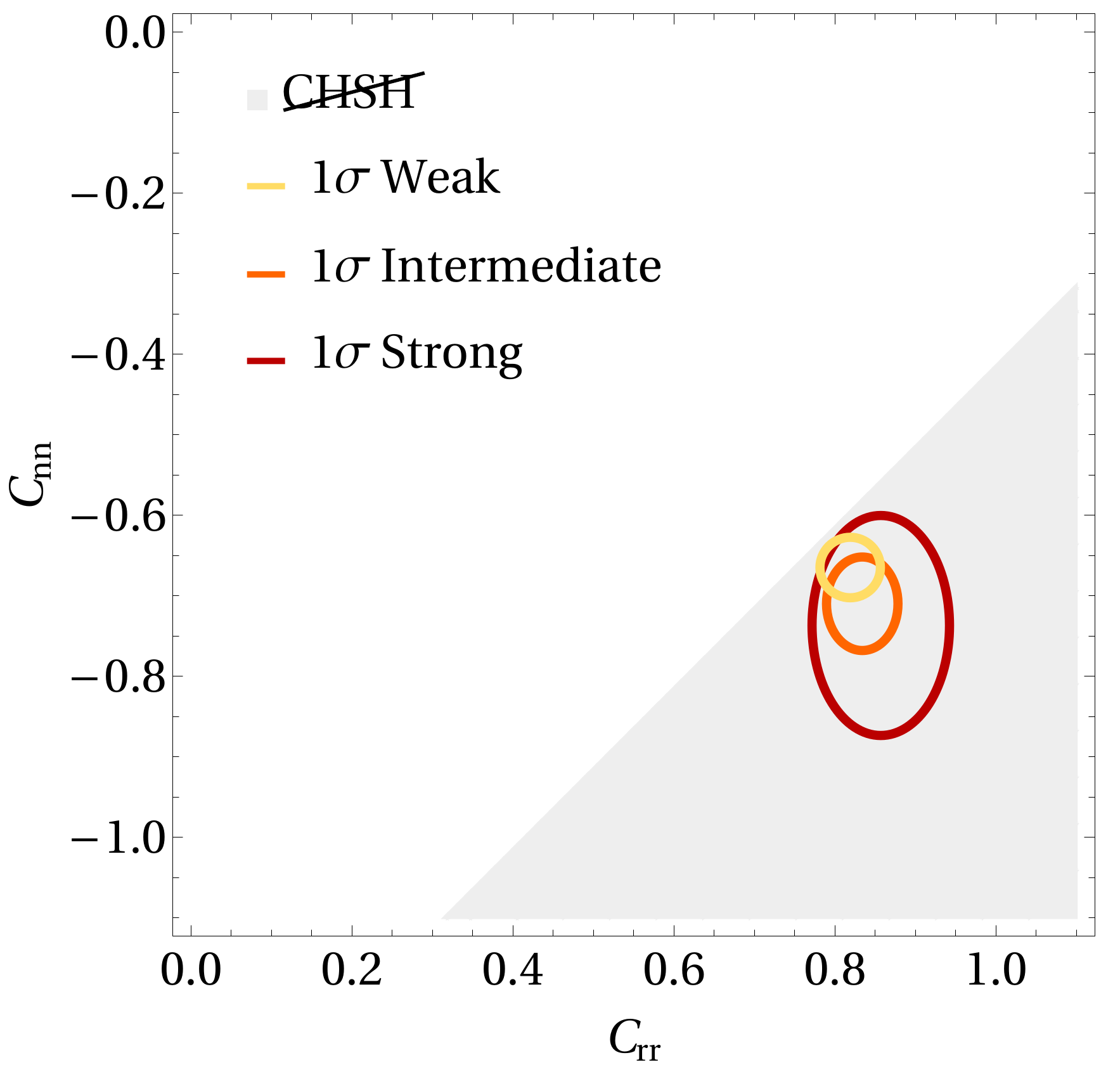}
	\captionsetup{width=\linewidth}
	\caption{Representation of results of Table \ref{table:chsh_results} in the $C_{rr} - C_{nn}$ plane. Shaded band: region where CHSH inequality is violated. Ellipses: $1 \sigma$ contour estimations for the value of $C_{rr}$ and $C_{nn}$ after the HL-LHC Run for the three selections. Stronger cuts move the central value further into the non--classical region, yet widen the uncertainties. It is expected \cite{cms3} that different entries of the $C$ matrix have different statistical uncertainties at the unfolded level, up to a relative factor of $\sim 2$. }
	\label{fig:results}
\end{figure}

\begin{figure}[h]
 \centering
  	\vspace{5mm}
 \includegraphics[width=\linewidth]{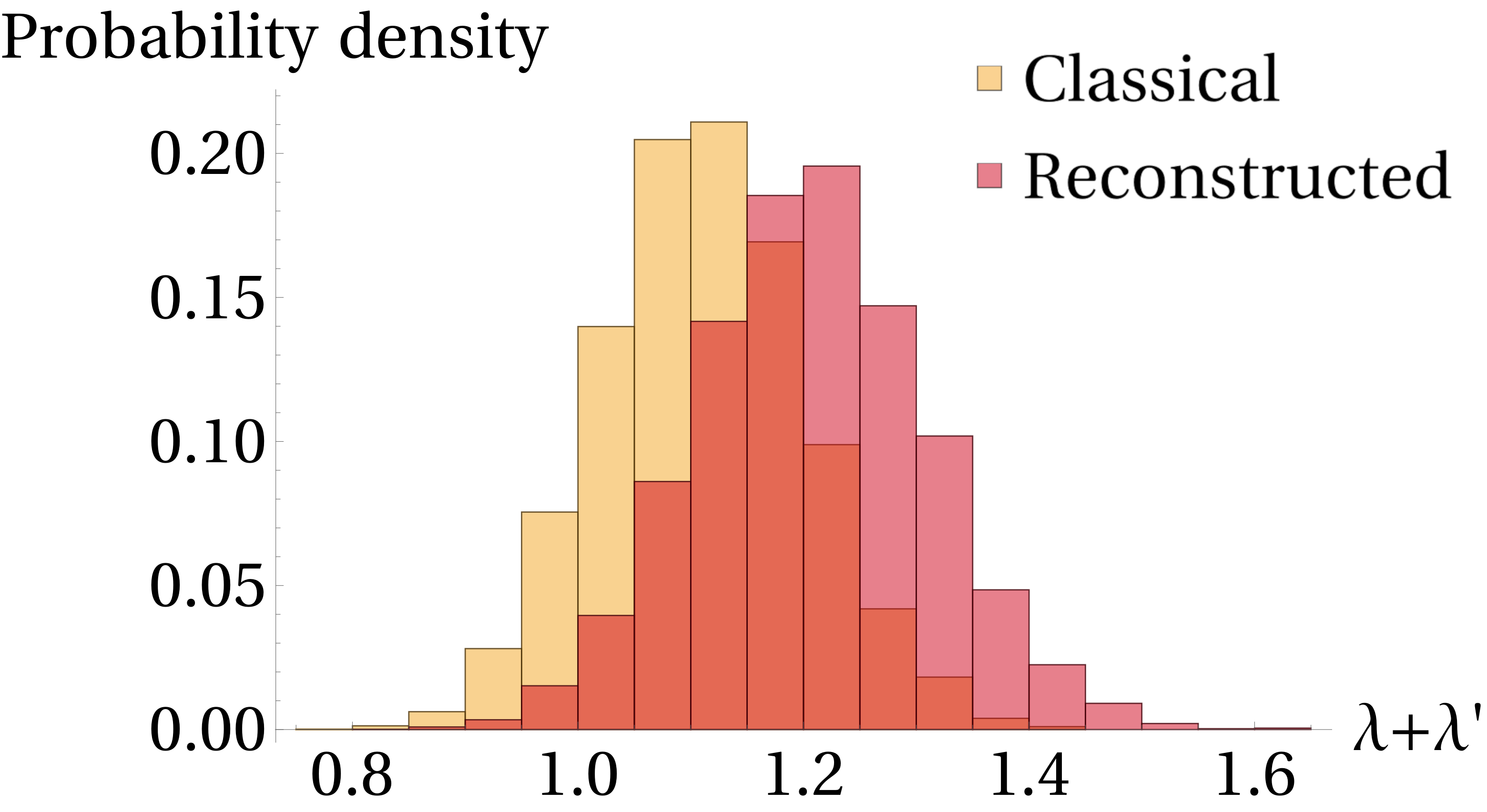}
	\captionsetup{width=\linewidth}
	\caption{Distribution of $\lambda+\lambda'$ following from reconstructed events and of $(\lambda+\lambda')_{\text{classical}}$ used in the hypothesis test for $\lambda+\lambda' > 1$, using the weak selection cuts described in the text.}
	\label{fig:hyptest}
\end{figure}

Table \ref{table:chsh_results} collects results for the fixed choice of axes of Eq.~\eqref{axis}. We find that the improvement given by the optimization is not enough to overcome the increase in systematic uncertainty noted in Sec. \ref{secsix}, and the overall performance of this method is worse than just using fixed axes. Finally, table \ref{table:chsh_results_eig} shows results for the hypothesis test using $\lambda + \lambda'$. Figure \ref{fig:hyptest} shows the distribution of $\lambda+\lambda'$ and $(\lambda+\lambda')_{\text{classical}}$ used for the hypothesis test with the weak selection cuts. We find that LHC Run 2 + Run 3 statistics are not sufficient for a conclusive measure. In order to provide an estimate for the upcoming High--Luminosity Run (HL-LHC), we estimate statistical uncertainties running all our analyses on $3 \, \text{ab}^{-1}$ of simulated luminosity. Results are shown in Figure \ref{fig:results}. The statistical significance for a violation of the CHSH inequality in Eq.~\eqref{chshhigh} becomes of order $\sim 2 \,\sigma$, regardless of the specific strategy or observable used.

	\begin{table*}
	\begin{center}
		\begin{tabular}{||l c | c c c c ||} \hline
		 High-$p_T$ & & \multicolumn{3}{c}{CHSH on fixed axes, $\sqrt{2} | - C_{rr} + C_{nn} | $} & Significance for $> 2$ \\
 			 Selection & Cross section & Parton--level & Reconstructed [$350 \; \text{fb}^{-1}$] & Reconstructed [$3 \; \text{ab}^{-1}$] & [$3 \, \text{ab}^{-1}$] \\ \hline
			{Weak} & $0.19 \, \text{pb}$ & $2.10$ & $2.12 \pm 0.17$ & $ \pm 0.06$ & $1.7 \, \sigma$ \\
			{Intermediate} & $0.10 \, \text{pb}$ & $2.18$ & $2.20 \pm 0.30$ & $ \pm 0.10$ & $1.8 \, \sigma$ \\
 			{Strong} & $0.06\, \text{pb}$ & $2.25$ & $2.30 \pm 0.76$ & $ \pm 0.26$ & $1.0 \, \sigma$ \\
 			\hline
		\end{tabular}
		\captionsetup{width=\linewidth}

		\caption{Results for the left-hand side of the CHSH inequality evaluated on the fixed axes in Eq.~\eqref{axis} for different final state selections (details in the text). For each selection an estimate of the cross section at $13 \, \text{TeV}$ is reported. The quoted central values are the average of several (14) independent simulated experiments using $350 \; \text{fb}^{-1}$ of simulated luminosity each. Uncertainties are statistical only, coming from the unfolding of $350 \; \text{fb}^{-1}$ (LHC Run 2 + 3) and $3 \, \text{ab}^{-1}$ of simulated luminosity.}
		\label{table:chsh_results}
	\end{center}
	\end{table*}

	\begin{table*}

	\begin{center}
		\begin{tabular}{|| l | c c||} \hline
		 High-$p_T$ & $\lambda + \lambda'$ & Significance for $> 1$ \\
 			 Selection & {Parton--level} & [$3 \, \text{ab}^{-1}$] \\ \hline
			{Weak} & $1.12$ & $1.9 \, \sigma$ \\
			{Intermediate} & $1.20$ & $2.1 \, \sigma$ \\
 			{Strong} & $1.30$ & $1.3 \, \sigma$ \\
 			\hline
		\end{tabular}
		\captionsetup{width=\linewidth}
		\caption{Replica of Table \ref{table:chsh_results}, showing results for the statistical significance of rejecting the "classical" hypothesis $\lambda + \lambda' = 1$ from $3 \; \text{ab}^{-1}$ of simulated luminosity (HL-LHC).}
		\label{table:chsh_results_eig}
	\end{center}
	\end{table*}	

 \section{Loopholes} 
 \label{ttloophole}
\begin{figure}
 \centering
 \includegraphics[width=.93\linewidth]{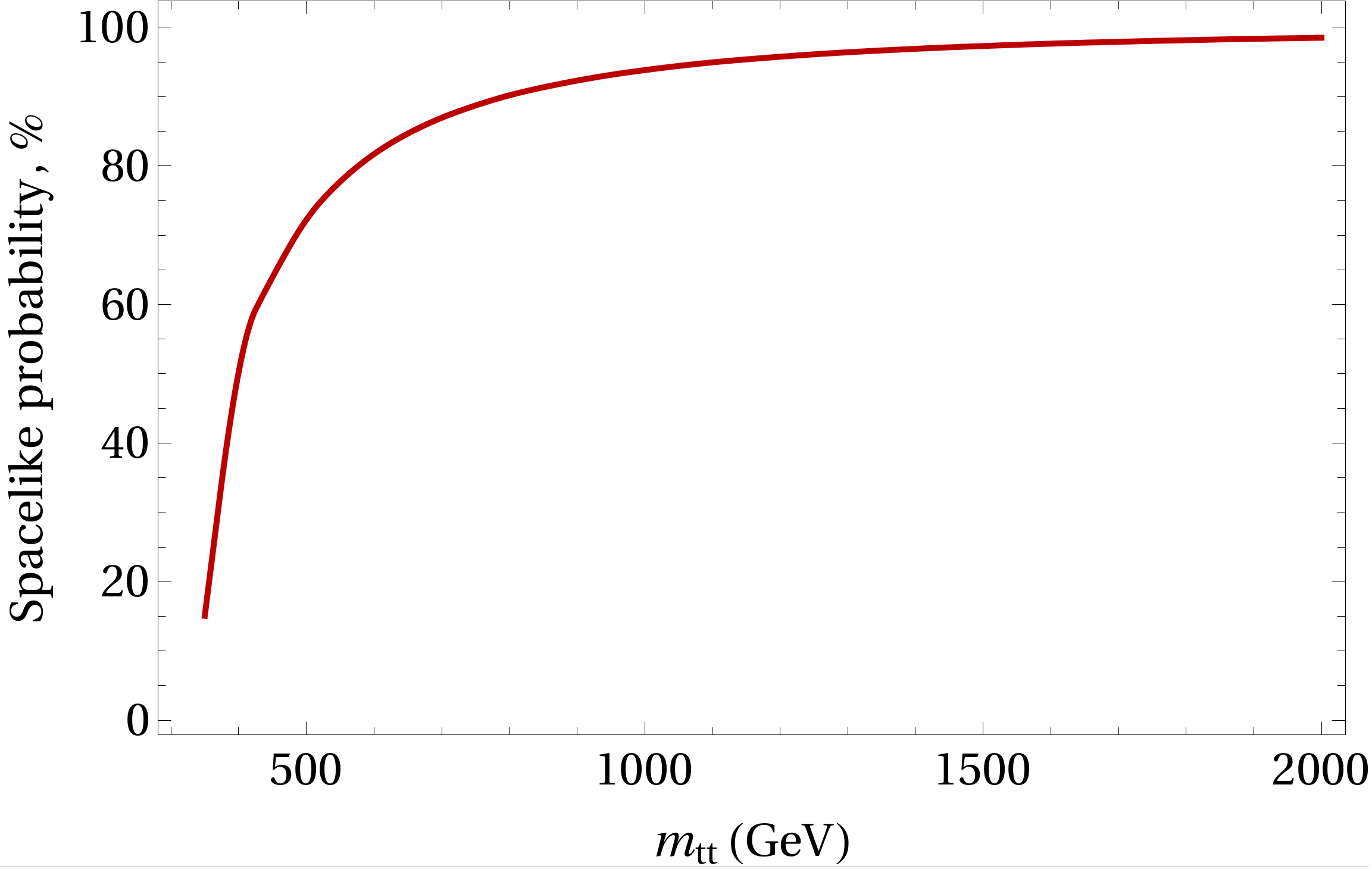}
	\captionsetup{width=\linewidth}
	\caption{Fraction of $t$ and $\bar t$ decays that are space-like separated as a function of $m_{t \bar t}$.}
	\label{fig:spacelike}
\end{figure}

When performing a Bell experiment the possible existence of loopholes has to be assessed. 
First, Bell experiments require outside intervention to choose {\it freely}, {\it i.e.} unknown to the system itself, what to measure. This can be achieved, for example, by mechanisms that randomly choose the orientations of the measurement axes. In the case of a $t \bar t$ system, no outside intervention is possible. However, one can argue that a random choice of axes is realised by the direction of the final state leptons in the top quark decays. 
 Second, the quantum measurements on the spin state of the $t \bar t$ pair, which take place when the top quarks decay leptonically, happen at very short distances. This is in contrast with the typical Bell experiment setup, which features macroscopic distances, and relates events that are always casually disconnected. In the case of the $t \bar t$ system, one can establish the casual independence of the decays only at the statistical level. In Figure \ref{fig:spacelike} we plot a Monte Carlo evaluation of the probability of space-like separated decays as a function of the pair invariant mass $m_{t \bar t}$. Close to threshold, most of the top pairs decay within each other's light-cone, while more than $90\%$ of $t \bar t$ pairs decay when they are space--like separated for $m_{t \bar t} > 800 \, \text{GeV}$.
Third, even assuming $b$--tagging and lepton identification were perfect, one can only observe $p p \to \ell^- \, \ell^+ \, b \, \bar b \, + E^{\text{miss}}$. In fact, top quarks might even not be present in a given event. However, within our simulations, we have verified that after reconstruction the large majority of the events selected can be attributed to $t \bar t$ pair production. 
Fourth, only a small fraction of events is usable for the analysis, and one has to assume the events that are recorded provide an unbiased representation of the bulk.

\section{Conclusions}
\label{conclusions}

We have presented new detailed studies for the detection of entanglement and the violation of Bell inequalities using spin-correlation observables in top quark pairs at the LHC. The main motivation for such a measurement is the possibility of performing a first TeV--scale Bell experiment, opening new prospects for high--energy precision tests of QM and of the SM. We have identified a unique set of observables that are sensitive to either the presence of entanglement or to a violation of a CHSH inequality. Our results indicate that the detection of entanglement will be straightforward, in agreement with Ref. \cite{afik2020}, and barring unexpected effects from systematic uncertainties, the LHC Run 2 dataset should be enough to reach a $5\sigma$ statistical significance. 

On the other hand, assessing the violation of Bell inequalities is much more challenging: sufficiently strong correlations are found only for top quarks at very high-$p_T$, thereby drastically reducing the available statistics. By considering only dileptonic final states and ignoring possibly relevant systematic uncertainties, whose evaluation goes beyond the scope of this study, we find the statistical significance for a violation to be of order $2 \sigma$ at the end of the High--Luminosity Run. The need to compare with theoretical expectations at unfolded level introduces a significant degradation of the naively estimated sensitivity just based on the share statistics, consistently with the current sensitivity of spin-correlations measurements at the LHC. 

The analysis strategy presented here is robust and can be directly implemented by the experimental collaborations. As a cross-check, we also directly employed the test statistics proposed in Ref. \cite{fabbrichesi2021} and obtained results which are consistent with our own method.

Barring the obvious benefits of an increased collider energy/luminosity, further studies to improve the prospects could be envisaged. For example, one could consider whether the limited statistics could be improved by including final states where only one top quark decays leptonically.

\section{Acknowledgements}

We are thankful to Eleni Vryonidou and Marco Zaro for comments on the manuscript and to Alan Barr, Stefano Forte, Rikkert Frederix and Marino Romano for discussions. 
C. S. is supported by the European Union’s Horizon 2020 research and innovation programme under the EFT4NP project (grant agreement no. 949451). F. M. has received funding from the European Union’s Horizon 2020 research and innovation programme as part of the Marie Skłodowska-Curie Innovative Training Network MCnetITN3 (grant agreement no. 722104) and by the F.R.S.-FNRS under the ‘Excellence of Science‘ EOS be.h project n. 30820817. 

\newpage
\bibliography{biblio}

\begin{thebibliography}{25}
\providecommand{\natexlab}[1]{#1}
\providecommand{\url}[1]{\texttt{#1}}
\expandafter\ifx\csname urlstyle\endcsname\relax
  \providecommand{\doi}[1]{doi: #1}\else
  \providecommand{\doi}{doi: \begingroup \urlstyle{rm}\Url}\fi

\bibitem[Bell(1964)]{bell}
J.~S. Bell.
\newblock On the {E}instein {P}odolsky {R}osen paradox.
\newblock \emph{Physics Physique Fizika}, 1\penalty0 (3):\penalty0 195--200,
  1964.
\newblock \doi{10.1103/PhysicsPhysiqueFizika.1.195}.

\bibitem[Afik and de~Nova(2021)]{afik2020}
Y.~Afik and J.R.M. de~Nova.
\newblock Quantum information and entanglement with top quarks at the {LHC}.
\newblock \emph{The European Physical Journal Plus}, 136, 2021.
\newblock \doi{10.1140/epjp/s13360-021-01902-1}.
\newblock \url{http://arxiv.org/abs/2003.02280}.

\bibitem[Fabbrichesi et~al.(2021)Fabbrichesi, Floreanini, and
  Panizzo]{fabbrichesi2021}
M.~Fabbrichesi, R.~Floreanini, and G.~Panizzo.
\newblock Testing {B}ell inequalities at the {LHC} with top-quark pairs.
\newblock \emph{Physical Review Letters}, 127, 2021.
\newblock \doi{10.1103/PhysRevLett.127.161801}.
\newblock \url{http://arxiv.org/abs/2102.11883}.

\bibitem[Takubo et~al.(2021)]{takubo2021}
Y.~Takubo et~al.
\newblock On the feasibility of bell inequality violation at atlas with flavor
  entanglement of b0-b0 pairs.
\newblock 2021.
\newblock \url{http://arxiv.org/abs/2106.07399}.

\bibitem[Barr(2021)]{barr2021}
A.~Barr.
\newblock Testing {B}ell inequalities in {H}iggs boson decays.
\newblock 2021.
\newblock \url{http://arxiv.org/abs/2106.01377}.

\bibitem[Clauser et~al.(1969)Clauser, M.A., Shimony, and Holt]{chsh}
J.F. Clauser, Horne M.A., A.~Shimony, and R.A. Holt.
\newblock Proposed experiment to test local hidden-variable theories.
\newblock \emph{Physical Review Letters}, 23\penalty0 (15):\penalty0 880--884,
  1969.
\newblock \doi{10.1103/PhysRevLett.23.880}.

\bibitem[Werner(1989)]{werner1989}
R.~F. Werner.
\newblock Quantum states with {E}instein-{P}odolsky-{R}osen correlations
  admitting a hidden-variable model.
\newblock \emph{Physical Review A}, 40, 1989.
\newblock \doi{10.1103/PhysRevA.40.4277}.

\bibitem[Peres(1996)]{peres1996}
A.~Peres.
\newblock Separability criterion for density matrices.
\newblock \emph{Physical Review Letters}, 77, 1996.
\newblock \doi{10.1103/PhysRevLett.77.1413}.
\newblock \url{http://arxiv.org/abs/quant-ph/9604005}.

\bibitem[Horodecki et~al.(1995)Horodecki, Horodecki, and
  Horodecki]{horodecki1995}
R.~Horodecki, P.~Horodecki, and M.~Horodecki.
\newblock Violating {B}ell inequality by mixed spin-$\frac{1}{2}$ states:
  necessary and sufficient condition.
\newblock \emph{Physics Letters A}, 200, 1995.
\newblock \doi{10.1016/0375-9601(95)00214-N}.

\bibitem[Bernreuther et~al.(2001)Bernreuther, Brandenburg, Si, and
  Uwer]{bernreuther2001}
W.~Bernreuther, A.~Brandenburg, Z.~G. Si, and P.~Uwer.
\newblock {Top quark spin correlations at hadron colliders: Predictions at
  next-to-leading order {QCD}}.
\newblock \emph{Physical Review Letters}, 87:\penalty0 242002, 2001.
\newblock \doi{10.1103/PhysRevLett.87.242002}.
\newblock \url{http://arxiv.org/abs/hep-ph/0107086}.

\bibitem[Frederix et~al.(2021)Frederix, Tsinikos, and Vitos]{frederix2021}
R.~Frederix, I.~Tsinikos, and T.~Vitos.
\newblock {Probing the spin correlations of $t{\bar{t}}$ production at {NLO}
  {QCD}+{EW}}.
\newblock \emph{The European Physical Journal C}, 81, 2021.
\newblock \doi{10.1140/epjc/s10052-021-09612-9}.
\newblock \url{http://arxiv.org/abs/2105.11478}.

\bibitem[Bernreuther et~al.(2004)Bernreuther, Brandenburg, Si, and
  Uwer]{bernreuther2004}
W.~Bernreuther, A.~Brandenburg, Z.G. Si, and P.~Uwer.
\newblock Top quark pair production and decay at hadron colliders.
\newblock \emph{Nuclear Physics B}, 690\penalty0 (1-2):\penalty0 81--137, 2004.
\newblock \doi{10.1016/j.nuclphysb.2004.04.019}.
\newblock \url{http://arxiv.org/abs/hep-ph/0403035}.

\bibitem[Chen et~al.(2013)Chen, Nakaguchi, and Komamiya]{chen2013}
S.~Chen, Y.~Nakaguchi, and S.~Komamiya.
\newblock Testing {B}ell's inequality using charmonium decays.
\newblock \emph{Progress of Theoretical and Experimental Physics},
  2013\penalty0 (6), 2013.
\newblock \doi{10.1093/ptep/ptt032}.
\newblock \url{http://arxiv.org/abs/1302.6438}.

\bibitem[Furedi and Komlos(1981)]{furedi1981}
Z.~Furedi and J.~Komlos.
\newblock The eigenvalues of random symmetric matrices.
\newblock \emph{Combinatorica}, 1, 1981.

\bibitem[Alwall et~al.(2014)]{madgraph}
J.~Alwall et~al.
\newblock The automated computation of tree-level and next-to-leading order
  differential cross sections, and their matching to parton shower simulations.
\newblock \emph{Journal of High Energy Physics}, 79, 2014.
\newblock \doi{10.1007/JHEP07(2014)079}.
\newblock \url{http://arxiv.org/abs/1405.0301}.

\bibitem[Group(2020)]{pdg}
Particle~Data Group.
\newblock Review of particle physics.
\newblock \emph{Progress of Theoretical and Experimental Physics},
  2020\penalty0 (8), 2020.
\newblock \doi{10.1093/ptep/ptaa104}.
\newblock \url{http://pdg.lbl.gov}.

\bibitem[Sjöstrand et~al.(2015)]{pythia8}
T.~Sjöstrand et~al.
\newblock An introduction to {PYTHIA} 8.2.
\newblock \emph{Computer Physics Communications}, 191, 2015.
\newblock \doi{10.1016/j.cpc.2015.01.024}.
\newblock \url{http://arxiv.org/abs/1410.3012}.

\bibitem[Collaboration(2014)]{delphes}
{DELPHES} Collaboration.
\newblock {DELPHES} 3: a modular framework for fast simulation of a generic
  collider experiment.
\newblock \emph{Journal of High Energy Physics}, 57, 2014.
\newblock \doi{10.1007/JHEP02(2014)057}.
\newblock \url{http://arxiv.org/abs/1307.6346}.

\bibitem[Collaboration(2018)]{atlas5}
{ATLAS} Collaboration.
\newblock Probing the quantum interference between singly and doubly resonant
  top-quark production in $pp$ collisions at $\sqrt{s} = 13$ {T}e{V} with the
  {ATLAS} detector.
\newblock \emph{Physical Review Letters}, 121, 2018.
\newblock \doi{10.1103/PhysRevLett.121.152002}.
\newblock \url{http://arxiv.org/abs/1806.04667}.

\bibitem[Collaboration(2019)]{cms3}
{CMS} Collaboration.
\newblock Measurement of the top quark polarization and $t \bar t$ spin
  correlations using dilepton final states in proton-proton collisions at
  $\sqrt{s}$ = 13 {T}e{V}.
\newblock \emph{Physical Review D}, 100, 2019.
\newblock \doi{10.1103/PhysRevD.100.072002}.
\newblock \url{http://arxiv.org/abs/1907.03729}.

\bibitem[Severi(2021)]{iome}
C.~Severi.
\newblock Bell inequalities with top quark pairs with the {ATLAS} detector at
  the {LHC}.
\newblock 2021.
\newblock \url{http://amslaurea.unibo.it/23535/}.

\bibitem[D'Agostini(2010)]{iterativebayesian}
G.~D'Agostini.
\newblock Improved iterative {B}ayesian unfolding.
\newblock 2010.
\newblock \url{http://arxiv.org/abs/1010.0632}.

\bibitem[Adye et~al.(2021)]{roounfold}
T.~Adye et~al.
\newblock Roounfold gitlab repository.
\newblock 2021.
\newblock \url{https://gitlab.cern.ch/RooUnfold/RooUnfold}.

\bibitem[Hoecker and Kartvelishvili(1996)]{hocker1995}
A.~Hoecker and V.~Kartvelishvili.
\newblock {SVD} approach to data unfolding.
\newblock \emph{Nuclear Instruments and Methods in Physics Research A}, 372,
  1996.
\newblock \doi{10.1016/0168-9002(95)01478-0}.
\newblock \url{http://arxiv.org/abs/hep-ph/9509307}.

\bibitem[Artoisenet et~al.(2013)]{madspin}
P.~Artoisenet et~al.
\newblock Automatic spin-entangled decays of heavy resonances in {M}onte
  {C}arlo simulations.
\newblock \emph{Journal of High Energy Physics}, 3, 2013.
\newblock \doi{10.1007/JHEP03(2013)015}.
\newblock \url{http://arxiv.org/abs/1212.3460}.

\end{thebibliography}

\end{document}